\documentstyle[11pt]{article}

\begin{document}
\begin{center}
\textbf{THE ENERGY CONSERVATION LAW 
IN CLASSICAL ELECTRODYNAMICS }
\end{center}

\begin{center}
E. G. Bessonov
\end{center}

\begin{center}
\textit{Lebedev Physical Institute RAS, 119991, Leninsky Prospect 53, Moscow, Russia}
\end{center}

\begin{center}
C. A. Brau
\end{center}

\begin{center}
\textit{Vanderbilt University, Nashville, TN 37235, USA }
\end{center}

\begin {abstract} A logical error in the usual derivation of the energy
conservation law is analyzed, and a way to avoid the error is presented.
\end {abstract}

In earlier papers \cite{Bessonov:2000}, \cite{Bessonov:2002} we
identified a logical error adopted by repetition in textbooks on
classical electrodynamics when the laws of conservation of energy and
momentum are derived for a system consisting of electromagnetic fields
and charged particles. In this paper we analyze the derivation of the
energy conservation law, discuss the origin of the logical error made
in this derivation, and present a way to avoid the error.

We start by reviewing briefly the usual derivation of the energy
conservation law in classical electrodynamics. By combining the Maxwell
equations in the usual way and integrating over a volume $V$ bounded by
the surface $A$, using the divergence theorem, we obtain Poynting's
theorem \cite{Landau:1975}, \cite{Jackson:1998}

\begin{equation}
\label{eq1}
\frac{d}{dt}\int\limits_V {wdV} + \oint\limits_A {{\rm {\bf S}} \cdot {\rm
{\bf \hat {n}}}dA} + \int\limits_V {{\rm {\bf J}} \cdot {\rm {\bf E}}dV} =
0
\end{equation}

In this expression, the quantity

\begin{equation}
\label{eq2}
w = \frac{\varepsilon _0 }{2}E^2 + \frac{1}{2\mu _0 }B^2
\end{equation}

\noindent
is the energy density of the electromagnetic field (SI units are used
throughout), ${\rm {\bf E}}$ the electric field, ${\rm {\bf B}}$ the
magnetic induction, ${\rm {\bf J}} = \rho {\rm {\bf v}}$ the current
density, $\rho $ the charge density, ${\rm {\bf v}}$ the velocity of
the charge in volume element $dV, {\rm {\bf S}} = {\rm {\bf E}}\times
{\rm {\bf B}} / \mu _0 $ the Pointing vector, ${\rm {\bf \hat {n}}}$
the unit vector normal to the surface $A$, and $dA$ the element of area
of the bounding surface.

Up to this point there are no inconsistencies, at least for continuous
distributions of charge and electromagnetic fields that are free of
singularities. The problems arise when Poynting's theorem is
generalized to include point charges, for in this case the total energy
in the fields (the first term in (\ref{eq1})) diverges and the field
${\rm {\bf E}}$ at the position of the charge (in the last term in
(\ref{eq1})) is not defined. The problems become worse when we
generalize (\ref{eq1}) to construct a global law for the conservation
of electromagnetic and mechanical energy, for in this case the
self-forces of the electromagnetic fields of a particle on itself
contribute to the inertia of the particle and to the radiation reaction
on the particle \cite{Landau:1975}, \cite{Jackson:1998}. We
ordinarily include the inertial effect of the self-fields in the
observed mass of the particle, and therefore double-count it when we
add the kinetic energy of the particles to the energy of the fields to
find the total energy in the system. The radiation reaction has its
problems as well: it causes unphysical motions of the particle. Thus,
the electrodynamics of point charges is fraught with contradictions.
They persist in quantum electrodynamics.

To simplify the following discussion, we extend the volume $V$ to
include all space and consider a closed system of particles, so there
are no ``external'' fields coming in from infinity. External fields can
be regarded as the fields from other particles that we include in the
closed system. If the fields vanish sufficiently rapidly at infinity,
the integral of the poynting vector over the surface $A$ vanishes and
we are left with

\begin{equation}
\label{eq3}
\frac{dW}{dt} + \int {{\rm {\bf J}} \cdot {\rm {\bf E}}dV} = 0,
\end{equation}

\noindent
where

\begin{equation}
\label{eq4}
W = \int {wdV} .
\end{equation}

When the current density arises from a set of point charges $q_i $ at
positions ${\rm {\bf r}}_i $, the current density ${\rm {\bf J}}$ may
be expressed in the form

\begin{equation}
\label{eq5}
{\rm {\bf J}} = \sum\limits_i {q_i {\rm {\bf v}}_i \delta \left( {{\rm {\bf
r}} - {\rm {\bf r}}_i } \right)} ,
\end{equation}

\noindent
where ${\rm {\bf v}}_i = d{\rm {\bf r}}_i / dt$ is the velocity of the
$i^{\mbox{th}}$ particle. When we substitute this into (\ref{eq3}) and
integrate over all space, the conservation law becomes

\begin{equation}
\label{eq6}
\frac{dW}{dt} + \sum\limits_i {q_i {\rm {\bf E}}\left( {{\rm {\bf r}}_i }
\right) \cdot {\rm {\bf v}}_i } = 0.
\end{equation}

But the rate at which work is done on the $i^{\mbox{th}}$ particle is
just the rate of increase of the energy $e_i $ of the $i^{\mbox{th}}$
particle,

\begin{equation}
\label{eq7}
\frac{de_i }{dt} = q_i {\rm {\bf E}}\left( {{\rm {\bf r}}_i } \right) \cdot
{\rm {\bf v}}_i .
\end{equation}

If we substitute this into (\ref{eq6}), we get the global conservation
law

\begin{equation}
\label{eq8}
\frac{d}{dt}\left( {W + E} \right) = 0,
\end{equation}

\noindent
where

\begin{equation}
\label{eq9}
E = \sum\limits_i {e_i } .
\end{equation}

\noindent
is the total energy of the particles.

However, for point charges, this derivation has three problems. In the
first place, the electromagnetic energy density $w$ in the fields of
the particles diverges at the positions of the particles. In the second
place, the field ${\rm {\bf E}}\left( {{\rm {\bf r}}_i } \right)$
includes the self-fields ${\rm {\bf E}}_i \left( {{\rm {\bf r}}_i }
\right)$. It is not defined at the position of the i$^{th}$ particle.
Typically, ${\rm {\bf E}}\left( {{\rm {\bf r}}_i } \right)$ is
restricted to the field of the other particles, which is well behaved
at ${\rm {\bf r}}_i $, and in the derivation of the energy conservation
law, the self-fields ${\rm {\bf E}}_i \left( {{\rm {\bf r}}_i }
\right)$ of the particle (the inertial and the radiation reaction) are
ignored \cite{Landau:1975}, \cite{Jackson:1998}. This is an annoying
error. In the third place, the energy of the Coulomb field surrounding
a particle is generally included as part of the mechanical energy of
the particle. That is, in the nonrelativistic limit

\begin{equation}
\label{eq10}
e_i = \textstyle{1 \over 2}m_i v_i^2 ,
\end{equation}

\noindent
where the observed mass $m_i $ includes both the ``bare mass'' of the
particle and the kinetic energy attributable to the
self-electromagnetic field of the particle. Since this electromagnetic
contribution is infinite, for a point charge, the bare mass of the
particle is assumed to be negative and infinite in a way that nearly
cancels out the electromagnetic contribution, leaving a finite observed
mass. This is called renormalization. Even if we set aside questions
about the validity of dealing with divergent quantities in this way, it
still means that we have counted the energy of the electromagnetic
fields of the particles twice in the conservation law (\ref{eq8}), once
in $W$ and once in $E$. Thus, we have introduced a logical
inconsistency, and the conservation law expressed by (\ref{eq7}) -
(\ref{eq9}) is incorrect as it stands. Unfortunately, this error
appears in almost every text on electrodynamics \cite{Landau:1975},
\cite{Jackson:1998}.  Conservation laws for the linear momentum and
angular momentum (or the 4-vector momentum) similar to (\ref{eq7}) -
(\ref{eq9}) can also be derived, but they have the same fundamental
problems. Moreover, the energy and momentum of the self-field do not
have the correct relativistic transformation properties.  This is the
so-called 4/3 problem, and it is resolved only by introducing the
so-called Poincar\'{e} stresses \cite{Poincar(c):1906}, \cite{
Rohrlich:1990}, \cite{Brau:2004} In the following we consider the
nonrelativistic case, since all the difficulties (including the 4/3
problem) are exhibited, and the relativistic effects are not essential
to the argument.

We begin with the last term in (\ref{eq6}) or (\ref{eq7}), and focus
our attention on the self-field contribution ${\rm {\bf E}}_i \left(
{{\rm {\bf r}}_i } \right)$ to ${\rm {\bf E}}\left( {{\rm {\bf r}}_i }
\right)$. To see how to handle the self-interaction, we consider the
motion of an extended charge distribution and then examine the limit
when the size of the distribution vanishes. Unfortunately, if we take
the limit in the usual way we obtain an equation of motion that has
nonphysical solutions. The details are described in the Appendix, but
for now it is enough to point out that the self-electromagnetic force
on a small, spherically symmetric, shell of charge can be represented
by the series

\begin{equation}
\label{eq11}
{\rm {\bf F}}_{i,(self)}^{(electromagnetic)} = q_i < {\rm {\bf E}}_i \left(
{{\rm {\bf r}}_i } \right) > = - \frac{q_i^2 }{6\pi \varepsilon _0
c^2a}\frac{d{\rm {\bf v}}_i }{dt} + \frac{q_i^2 }{6\pi \varepsilon _0
c^3}\frac{d^2{\rm {\bf v}}_i }{dt^2} + O\left( a \right),
\end{equation}

\noindent
in the nonrelativistic case, where $a$ is the radius of the charge
distribution and brackets <> denote averaging through the particle
volume.  In the nonrelativistic limit, which is valid in the particle
rest frame, we can ignore magnetic effects. If we overlook the fact
that the derivation is based on the concept of a rigid charge
distribution, which is impossible in relativistic kinematics, the
relativistic generalization of (\ref{eq11}) is straightforward. The
first term in (\ref{eq11}) is negative and proportional to the
acceleration. It resists the acceleration in just the way that the bare
mass does, so we can add it to the inertial term in the equation of
motion. It contributes an electromagnetic mass

\begin{equation}
\label{eq12}
m_{(dynamic)}^{(electromagnetic)} = \frac{q^2}{6\pi \varepsilon _0 c^2a},
\end{equation}

\noindent
which diverges as the particle shrinks to a point ($a \to 0)$. Note
that the rest mass that appears is $m_{(dynamic)}^{(electromagnetic)} =
\textstyle{4 \over 3}m_{(energetic)}^{(electromagnetic)} $, where
$m_{(energetic)}^{(electromagnetic)} c^2 = q^2 / 8\pi \varepsilon _0 a$
is the energy of the electrostatic field around a shell of charge of
radius $a$. This is called the 4/3 problem, and the difference is
attributable to the so-called Poincar\'{e} stresses that hold the
charged particle together \cite{Jackson:1998}, \cite{
Poincar(c):1906}. The next term in the expansion (\ref{eq11}) of the
self-force is called the radiation reaction. It remains finite as the
particle shrinks to a point, but is responsible for runaway solutions,
as described in the appendix. We ignore these problems in the following
discussion, and in fact they can be avoided by taking the limit in a
way that represents the self-force by a difference-differential
equation or integro-differential equation.  The remaining terms in the
series (\ref{eq11}) vanish in the limit as the particle shrinks to a
point.

By analogy we can represent the self-nonelectromagnetic force on the
same shell in the form

\begin{equation}
\label{eq13}
{\rm {\bf F}}_{i,(self)}^{(nonelectromagnetic)} = -
m_{i,(dynamic)}^{(nonelectromagnetic)} \frac{d{\rm {\bf v}}_i }{dt}.
\end{equation}

The equation (\ref{eq13}) is limited to one term (we took into account
that the fields of nonelectromagnetic origin are not emitted and hence
do not produce a radiation reaction).

If we postulate that the external force applied to the i$^{th}$
particle is equal to the sum of electromagnetic and nonelectromagnetic
self-forces and is of opposite direction, then, according to
(\ref{eq11}), (\ref{eq13}), the equation of motion can be presented in
the form accepted in the classical electrodynamics

\begin{equation} \label{eq14} m_i^{\left( {observed} \right)}
\frac{d{\rm {\bf v}}_i }{dt} = q_i {\rm {\bf E}}^{(other)} +
\frac{q_i^2 }{6\pi \varepsilon _0 c^3}\frac{d^2{\rm {\bf v}}_i }{dt^2}
+ 0(a), \end{equation}

\noindent
where $m_i^{\left( {observed} \right)} = m_{i,(dynamic)}^{
(electromagnetic)} + m_{i,(dynamic)}^{(nonelectromagnetic)} $ is the
observed mass, ${\rm {\bf E}}^{(other)} \left( {{\rm {\bf r}}_i }
\right) = \sum _{j \ne i} {{\rm {\bf E}}_j \left( {{\rm {\bf r}}_i }
\right)} $ the total electric field at the point ${\rm {\bf r}}_i $ due
to all the other particles $j \ne i$, and ${\rm {\bf E}}_j \left( {{\rm
{\bf r}}_i } \right)$ the field of particle $j$ at the point ${\rm {\bf
r}}_i $.

The equation (\ref{eq6}) includes the self-fields of the particle ${\rm
{\bf E}}_i({r{\bf}}_i)$. According to (\ref{eq11}) and (\ref{eq14}) the
value

\[
q_i {\rm {\bf E}}\left( {{\rm {\bf r}}_i } \right){\rm {\bf v}}_i = q_i <
{\rm {\bf E}}_i \left( {{\rm {\bf r}}_i } \right) > {\rm {\bf v}}_i + q_i
{\rm {\bf E}}^{(other)} \left( {{\rm {\bf r}}_i } \right){\rm {\bf v}}_i =
[m_i^{\left( {observed} \right)} - m_{i,(dynamic)}^{(electromagnetic)}
]\frac{d{\rm {\bf v}}_i }{dt}{\rm {\bf v}}_i
\]

\begin{equation}
\label{eq15}
 = K_i^{\left( {observed} \right)} - K_{i,(dynamic)}^{\left(
{electromagnetic} \right)}
\end{equation}

\noindent
where $K_i^{\left( {observed} \right)} = \textstyle{1 \over
2}m_i^{\left( {observed} \right)} v_i^2 $, is the ``observed'' kinetic
energy of the particles and\\ $K_{i,(dynamic)}^{\left(
{electromagnetic} \right)} = \textstyle{1 \over
2}m_{i,(dynamic)}^{\left( {electromagnetic} \right)} v_i^2 $ the
kinetic electromagnetic dynamic energy of the particles.

If we substitute (\ref{eq15}) into (\ref{eq6}), we obtain the
conservation law in the form

\begin{equation}
\label{eq16}
\frac{d}{dt}\left[ {W + K^{\left( {observed} \right)} -
K_{(dynamic)}^{(electromagnetic)} } \right] = 0,
\end{equation}

\noindent
where $K^{\left( {observed} \right)} = \sum\limits_i {K_i^{\left(
{observed} \right)} } $ is the total ``observed'' kinetic energy of the
particles,\\ $K_{(dynamic)}^{(elecromagnetic)} = \sum\limits_i
{K_{i,(dynamic)}^{\left( {electromagnetic} \right)} } $. Clearly,
$K_{i,(dynamic)}^{\left( {electromagnetic} \right)} $ is the energy
that has been double counted in (\ref{eq8}) \cite{Bessonov:2000}$,
$\cite{Bessonov:2002} .

The equation (\ref{eq16}) still contains divergent terms in $W$ and
$K_{(dynamic)}^{(elecromagnetic)} $ so it is impossible to use the
conservation law in this form. To deal with this, we can proceed in the
following manner. Since the field near the particle approaches the
Coulomb field of a homogeneously moving particle, we can write

\begin{equation}
\label{eq17}
{\rm {\bf E}}_i = {\rm {\bf E}}_i^{\left( {Coulomb} \right)} + {\rm {\bf
E}}_i^{\left( {remainder} \right)} ,
\end{equation}

\noindent
with a similar decomposition of the magnetic field. In the limit as the
size of the particle vanishes, we see that its electromagnetic energy
is

\begin{equation}
\label{eq18}
W_i = W_i^{\left( {Coulomb} \right)} + W_i^{\left( {remainder} \right)} ,
\end{equation}

\noindent
where the reminder energy of the i$^{th}$ particle is

\[
W_i^{\left( {remainder} \right)} = \frac{\varepsilon _0 }{2}
\int {\left(
2{\rm {\bf E}}_i^{\left( {Coulomb} \right)} \cdot {\rm {\bf E}}_i^{\left(
{remainder} \right)} + E_i^{{\left( {remainder} \right)} ^2} \right) dV
}
\]

\begin{equation}
\label{eq19}
+ \frac{1}{2\mu _0 }\int {\left( 2{\rm {\bf B}}_i^{\left( {Coulomb}
\right)} \cdot {\rm {\bf B}}_i^{\left( {remainder} \right)} +
B_i^{{\left( remainder \right)} ^2} \right)dV
},
\end{equation}

\noindent
which is finite.

In this case we can represent the total energy of the electromagnetic fields
in the form

\begin{equation}
\label{eq20}
W = \sum\nolimits_i {W_i } + \frac{\varepsilon _0 }{2}\int {\sum\limits_{i
\ne j} {{\rm {\bf E}}_i \cdot } {\rm {\bf E}}_j dV} \mbox{ }
 + \frac{1}{2\mu _0 }\int {\sum\limits_{i \ne j} {{\rm {\bf B}}_i \cdot {\rm
{\bf B}}_j } dV} = W^{Coulomb} + W^{remainder},
\end{equation}

\noindent
where the value $W^{(Coulomb)} = \sum\nolimits_i W _i^{(Coulomb)} $
corresponds to the energy of particles homogeneously moving with the
given velocity at the moment of observation and the electromagnetic
energy $W^{(remainder)} = \sum\nolimits_i W _i^{(remainder)} +
\frac{\varepsilon _0 }{2}\int {\sum\limits_{i \ne j} {{\rm {\bf E}}_i
\cdot } {\rm {\bf E}}_j dV} \mbox{ } + \frac{1}{2\mu _0 }\int
{\sum\limits_{i \ne j} {{\rm {\bf B}}_i \cdot {\rm {\bf B}}_j } dV} $
is the sum of the interaction energy of particles with electromagnetic
fields and the energy of emitted radiation.

The energy in the self-field of a homogeneously moving spherical shell
of a charge is

\[
W_i^{(Coulomb)} = \frac{\varepsilon _0 }{2}\int\limits_{r > a} {E_i^2 dV} +
\frac{1}{2\mu _0 }\int\limits_{r > a} {B_i^2 dV} = m_{i,(energetic)}^{\left(
{electromagnetic} \right)} c^2(1 + \beta _i^2 / 3) / \sqrt {1 - \beta _i^2 }
\vert _{\beta _i \ll 1}
\]

\begin{equation}
\label{eq21}
 = m_{i,(energetic)}^{\left( {electromagnetic} \right)} c^2 + \textstyle{5
\over 6}m_{i,(energetic)}^{\left( {electromagnetic} \right)} v_i^2 =
m_{i,(energetic)}^{\left( {electromagnetic} \right)} c^2 +
\frac{5}{3}K_{i,(energetic)}^{\left( {electromagnetic} \right)} ,
\end{equation}

\noindent
where $K_{i,(energetic)}^{(electromagnetic)} = \frac{1}{2}m_{i,(
energetic)} ^{\left( {electromagnetic} \right)} v_i^2 $ (see,
e.g., \cite{Butler:1969}).

If we substitute the total electromagnetic energy of the system $W =
W^{\left( {remainder} \right)} + W^{(Coulomb)}$ to (\ref{eq16}) and
neglect the derivative of the constant terms $m_{i,(energetic)}^{\left(
{electromagnetic} \right)} c^2$, the equation takes the form

\begin{equation}
\label{eq22}
\frac{d}{dt}\left[ {W^{\left( {remainder} \right)} + K^{\left( {observed}
\right)} + \left( {\frac{5}{3}K_{(energetic)}^{(electromagnetic)} -
K_{(dynamic)}^{(electromagnetic)} } \right)} \right] = 0,
\end{equation}

\noindent
where $W^{(Coulomb)} = \sum\limits_i {W_i^{(Coulomb)} } $ and
$K_{(energetic)}^{(electromagnetic)} = \sum\limits_i
{K_{i,(energetic)}^{\left( {electromagnetic} \right)} } $ is the kinetic
electromagnetic energy of particles homogeneously moving with the velocity
${\rm {\bf v}}_i$.

The difference term in parentheses of (\ref{eq22}) is $K_{i,(
energetic)} ^{\left( {electromagnetic} \right)} / 3 \ne 0$. We can
suppose that this difference term is attributable to the so-called
Poincar\'{e} stresses that hold the charged particle together, that if
the electromagnetic energy is transformed by a complicated law
(\ref{eq21}), the nonelectromagnetic energy is transformed by any
analogous law and compensate this difference. If we ignore it then this
leaves the conservation law

\begin{equation}
\label{eq23}
\frac{d}{dt}\left[ {W^{\left( {remainder} \right)} + K^{\left( {observed}
\right)}} \right] = 0.
\end{equation}

The conservation law (\ref{eq23}) can be expressed in the equivalent form

\begin{equation}
\label{eq24}
\frac{d}{dt}\left[ {W + K_{(energetic)}^{(nonelecromagnetic)} } \right] =
0,
\end{equation}

\noindent
where $K_{(energetic)}^{(nonelectromagnetic)} = K_{(dynamic)}^{
(nonelectromagnetic)} + \frac{1}{3}K_{(energetic)}
^{(nonelectromagnetic)} , K_{(dynamic)}^{(nonelectromagnetic)} =
\quad \sum\limits_i {\textstyle{1 \over 2}m_{i,(dynamic)}^{\left(
{nonelectromagnetic} \right)} v_i^2 } $.

In this form, the conservation law (\ref{eq24}) states explicitly that
the sum of the electromagnetic energy and the kinetic energy
attributable to the nonelectromagnetic energetic energy of particles is
a constant. It still contains divergent terms. We can extract from $W$
divergent terms corresponding to the accompanying electromagnetic
energy of particles, combine them with $K_{(energetic)}^{
(nonelectromagnetic)} $ and postulate that the obtained value is the
observed energy of particles. This can be called renormalization. It
leaves the conservation law (\ref{eq23}) in the same form but now the
observed kinetic energy is composed of electromagnetic and
nonelectromagnetic energies determined by masses of energetic origin.

Both the electromagnetic and nonelectromagnetic dynamic masses in the
equation of motion (\ref{eq14}) can be considered as coefficients at
the resistive terms. They have the dimensions of mass and their sum
agrees with the observable mass. The sum of electromagnetic and
nonelectromagnetic energetic masses of particles is another form of
presentation of the observable mass of particles based on their energy.

There is still the matter of the 4/3 problem and the energy due to the
Poincar\'{e} stresses. In the nonrelativistic theory we simply ignore
the divergent, constant terms in the field energy $W$, as discussed
above, so the problem disappears. In the relativistic case, however,
this energy is part of the rest energy of the particle. Unfortunately,
this energy is only 3/4 what we would expect from the mass that appears
in the momentum. The missing mass, as first pointed out by Poincar\'{e}, is
accounted for by considering the forces that hold the electric charge
distribution. In the simplest example, the momentum density of the
Poincar\'{e} stresses (the off-diagonal elements of the stress 4-tensor)
vanish in all coordinate systems, and the energy density is just
sufficient to make up the missing mass. More elaborate models of the
Poincar\'{e} stress tensor have both momentum and energy, but when the
Poincar\'{e} stresses are included in the symmetric stress tensor, a
covariant form of the conservation law can be derived
\cite{Jackson:1998}, \cite{Poincar(c):1906}.

To conclude, we have shown a logical inconsistency in the derivation of
the energy conservation law that appears in almost every text on
advanced electrodynamics. The conservation law expressed by (\ref{eq7})
- (\ref{eq9}) is incorrect as it stands and must be replaced by
(\ref{eq23}). Unfortunately we can not state that the correct formulae
for the energy conservation law was obtained without any hypotheses.
It is better to say that we proceed from the assumption that the energy
conservation law and Maxwell equations are valid in any case and put
limitations on the nonelectromagnetic fields.

              \begin {center}\bf  Appendix \end {center}

We start by reviewing very briefly the derivation of the rate of change
of momentum of the particle based on Abraham-Lorentz evaluation of the
self-force. In the nonrelativistic case, the equation of motion of a
particle in the external fields can be written in the form

\begin{equation}
\label{eq25}
m^{\left( {bare} \right)}\frac{d{\rm {\bf v}}}{dt} = q\left( {{\rm {\bf
E}}^{\left( {other} \right)} + {\rm {\bf v}}\times {\rm {\bf B}}^{\left(
{other} \right)}} \right) + {\rm {\bf F}}^{\left( {self} \right)},
\end{equation}

\noindent
where $m^{\left( {bare} \right)}$ is the mass of the bare particle
(without electromagnetic fields), ${\rm {\bf E}}^{\left( {other}
\right)}$ and ${\rm {\bf B}}^{\left( {other} \right)}$ are the electric
and magnetic fields of other particles in the system, and ${\rm {\bf
F}}^{\left( {self} \right)}$ is the electromagnetic force of the charge
distribution of the particle back on itself. In the nonrelativistic
limit we can ignore the effect of the self-magnetic field compared to
that of the electric field, and the effect of the self-electric field
can be expressed as an integral over the retarded self-field of the
particle of the form

\begin{equation}
\label{eq26}
{\rm {\bf F}}^{\left( {self} \right)}\left( t \right) = \frac{1}{4\pi
\varepsilon _0 }\int {d^3{\rm {\bf r}}} \rho \left( {{\rm {\bf r}},t}
\right)\int {d^3{\rm {\bf r}}'\left[ {\frac{1}{cR}\left( {{\rm {\bf \hat
{R}}}\frac{\partial \rho }{\partial t'} - \frac{1}{c}\frac{\partial {\rm
{\bf J}}}{\partial t'}} \right) + \frac{{\rm {\bf \hat {R}}}}{R^2}\rho }
\right]_{retarded} } ,
\end{equation}

\noindent
where

\begin{equation}
\label{eq27}
{\rm {\bf R}} = {\rm {\bf r}} - {\rm {\bf r}}',
\end{equation}

\noindent
and the retarded time is

\begin{equation}
\label{eq28}
t' = t_{retarded} = t - \frac{\left| {{\rm {\bf r}} - {\rm {\bf r}}'}
\right|}{c}.
\end{equation}

For a rigid charge distribution, the charge density at the retarded
time $t_{ret} $ is simply related to that at the present time $t$ by
the motion of the center of mass of the particle. If the charge
distribution is small, the retarded times are all close to the present
time and we can use a Taylor-series expansion to evaluate $\rho \left(
{{\rm {\bf r}}',t_{ret} } \right)$. For a rigid, spherical shell of
charge the result is

\begin{equation}
\label{eq29}
{\rm {\bf F}}^{\left( {self} \right)} = \frac{q^2}{12\pi \varepsilon _0
ca^3}\sum\limits_{n = 1}^\infty {\frac{\left( { - 1} \right)^n}{\left( n
\right)!}\left( {\frac{2a}{c}} \right)^n\frac{d^n{\rm {\bf v}}\left( t
\right)}{dt^n}} ,
\end{equation}

\noindent
where $q$ is the total charge, $a$ the radius, and ${\rm {\bf v}}\left(
t \right)$ the velocity of the charge \cite{Jackson:1998}, \cite
{Butler:1969}. The series can be summed, and the result is given by the
expression

\begin{equation}
\label{eq30}
{\rm {\bf F}}^{\left( {self} \right)}\left( t \right) = \frac{q^2}{12\pi
\varepsilon _0 ca^2}\left[ {{\rm {\bf v}}\left( {t - \frac{2a}{c}} \right) -
{\rm {\bf v}}\left( t \right)} \right].
\end{equation}

When this is substituted into the nonrelativistic equation of motion,
we get the difference-differential equation

\begin{equation}
\label{eq31}
m^{\left( {bare} \right)}\frac{d{\rm {\bf v}}}{dt} = q\left( {{\rm {\bf
E}}^{\left( {other} \right)} + {\rm {\bf v}}\times {\rm {\bf B}}^{\left(
{other} \right)}} \right) + \frac{q^2}{12\pi \varepsilon _0 a^2}\left[ {{\rm
{\bf v}}\left( {t - \frac{2a}{c}} \right) - {\rm {\bf v}}\left( t \right)}
\right].
\end{equation}

This is called the Page-Somerfeld equation of motion
\cite{Somerfeld:1904}, \cite{Page:1918}. Its relativistic
generalization is discussed by Caldirola \cite{Caldirola:1956}. For more
general charge distributions we get an integro-differential equation of
motion called the Markov equation \cite{Markov:1946}. Its relativistic
generalization is discussed by Brau \cite{Brau:2004}.

If instead of summing the series (\ref{eq29}) we take just the first
two terms, we get

\begin{equation}
\label{eq32}
{\rm {\bf F}}^{\left( {self} \right)} = - \frac{q^2}{6\pi \varepsilon _0
ac^2}\frac{d{\rm {\bf v}}}{dt} + \frac{q^2}{6\pi \varepsilon _0
c^3}\frac{d^2{\rm {\bf v}}}{dt^2}.
\end{equation}

If we substitute this into the equation of motion (\ref{eq25}) we get

\begin{equation}
\label{eq33}
m^{\left( {observed} \right)}\frac{d{\rm {\bf v}}}{dt} = q\left( {{\rm {\bf
E}}^{\left( {other} \right)} + {\rm {\bf v}}\times {\rm {\bf B}}^{\left(
{other} \right)}} \right) + \frac{q^2}{6\pi \varepsilon _0 a^2}\frac{d^2{\rm
{\bf v}}}{dt^2},
\end{equation}

\noindent
where the observed mass of the particle is given by (\ref{eq14}). The
result (\ref{eq33}) is called the Abraham-Lorentz equation of motion
\cite{Abraham:1905}, \cite{Lorentz:1952}. Its relativistic
generalization is discussed by Dirac \cite{Dirac:1938}. Unfortunately,
this equation admits runaway solutions, for if the external fields
vanish the equation of motion (\ref{eq33}) is satisfied by the solution

\begin{equation}
\label{eq34}
{\rm {\bf v}}\left( t \right) = {\rm {\bf v}}_0 e^{t / \tau },
\end{equation}

\noindent
where $\tau = 2a / 3c$ and ${\rm {\bf v}}_0 $ is a constant. That is,
in the absence of external fields the particle can start at rest and
accelerate without limit. Runaway solutions are avoided by the
Page-Somerfeld equation of motion and, under certain conditions, by the
Markov equation of motion.  However, these equations of motion admit
oscillatory solutions \cite{Bohm:1948}.

\end{document}